# Purcell and collection efficiency enhancement of single NV- center emission coupled to an asymmetric Tamm structure

Nitesh Singh and Rajesh V Nair[*]

Laboratory for Nano-scale Optics and Meta-materials (LaNOM)
Department of Physics, Indian Institute of Technology Ropar, Punjab 140001 India
Email: rvnair@iitrpr.ac.in

The resonant modes associated with engineered photonic structures of different spatial-dimension are essential to obtain bright on-demand single photon sources for quantum technologies. Negatively-charged nitrogen-vacancy (NV-) center in diamond is proposed as an excellent single photon source at room temperature. The possible optical readout of spin states in diamond makes it a good candidate for spin-photon interface. However, the poor light collection, feeble zero-phonon line, low emission rate, and the presence of broad phonon-induced emission limits the use of NV- center in quantum technologies. Here, we present a feasible, easy to fabricate, asymmetric Tamm structure coupled with single NV- center to enhance the emission rate of zero phonon line (ZPL) with better light collection efficiency. The asymmetric Tamm structure shows dual resonant mode with one of the modes coinciding with NV- ZPL wavelength of 640 nm. The quality factor ($Q$-factor) of the mode is 140 which is far better than the $Q$-factor of conventional Tamm structure and hence, better field intensity localization. We achieve four times Purcell enhancement along with an enhanced light collection efficiency of five times at 640 nm. The proposed structure is useful for generating bright single photons using NV- center and an efficient spin-photon interface.

# I. INTRODUCTION

On-demand generation of single photons with high emission rates and an efficient spin-photon interface using single quantum emitters requires trapping them in sub-wavelength photonic cavities [1-3]. The cavity modifies the electromagnetic environment around the emitter due to an increased local density of optical states (LDOS) at a specific wavelength (frequency) in comparison to vacuum. The LDOS specifies the total number of available electromagnetic modes per unit volume per unit frequency. The decay rate is directly related to LDOS through Fermi's golden rule and hence, for any reliable control on emission intensity and rate, LDOS engineering is an essential requirement [4]. Depending on the environmental changes around the emitter, the LDOS at the emission wavelength can be increased or decreased in a precise manner [5, 6]. This necessitates the development of a variety of sub-wavelength photonic cavity structures of different spatial dimensions embedded with quantum emitters. This includes photonic crystal cavities [6-8], plasmonic cavities [9], Mie-resonances [10], and resonant metasurfaces [11]. In such systems, the emission control is achieved using cavity mode with high quality factor ($Q$-factor) and minimum mode volume.

The recent developments in solid-state quantum emitters disseminate the negatively-charged nitrogen-vacancy (NV-) center in nanodiamond as a promising quantum emitter due to its excellent optical and spin properties at room temperature. The NV- center is formed by a substitutional nitrogen atom adjacent to a carbon vacancy in a diamond crystal, which finds applications in quantum communication, quantum sensing, magnetometry, and bio-markers [12-16]. The emission spectrum of NV- center consists of pure electronic transition at 640 nm called the zero phonon line (ZPL) assisted with broad phonon sideband (PSB) emission spanning up to 750 nm. The PSB transitions induce decoherence with a limited Debye-Waller factor of 3% emission at ZPL. Therefore, it is an ongoing effort to enhance emission intensity and rate at the ZPL with simultaneous PSB inhibition for efficient use of NV- center in quantum technologies [8, 14]. One approach is to cool the sample to ultra-low temperature to suppress phonon mediated transitions, but this results only in a slight increase in Debye-Waller factor to 4%, which calls for an alternate approach to enhance ZPL. Therefore, trapping NV- center to photonic cavity structures to achieve ZPL enhancement with simultaneous PSB suppression is proposed. Generally, the photonic crystal cavities could offer a high $Q$-factor but are restrained from offering low mode volume and involves complex fabrication procedures [17]. The photonic crystal cavities are fabricated on a monocrystalline diamond and were coupled to NV- center with cavity mode at its ZPL wavelength [8]. On the other hand, the plasmonic cavities have a low $Q$-factor of ~10 but possess a very small effective mode volume. This results in high Purcell factor and the coupling rate is enhanced to a large extent using plasmonic structures [18]. The coupling of NV centers and dye molecules with plasmonic cavities has been reported to achieve Purcell enhancement [19, 20]. However, the fabrication of plasmonic structures and then coupling of NVs with them is always a challenge in-addition to the inherent losses associated with plasmonic materials. Hence, the search for low-cost and easy to fabricate photonic cavity structures with moderate $Q$-values is a contemporary topic of research. The dielectric, metallic, and metal-dielectric layered structures have gained extensive attention due to their capability to enhance light-matter interactions. Hyperbolic metamaterials are metal-dielectric systems having hyperbolic dispersion which are used to obtain Purcell enhancement of ~250 at 650 nm [21]. To achieve maximum coupling between the emitter and cavity mode, the emitter must be positioned at a site of maximum localized field intensity inside the cavity. The cavity mode coupling with emitter ensures high emission rate and extraction efficiency [22]. The other interesting layered cavity structure is the optical Tamm structure which possesses surface states with moderate $Q$-values. The Tamm structure consists of a distributed Bragg reflector (DBR), spacer layer, and metal layer, respectively, forming a metal-dielectric layered structure. The structure is characterized by a Tamm mode, which originates due to the total reflection induced by the DBR and the metal film such that the mode is localized within spacer layer, termed Tamm plasmon resonance (TPR). The TPR have an in-plane wavevector smaller than the light wavevector in vacuum, allowing direct TPR excitation in contrast to surface plasmon polaritons. The TPR can be excited using both transverse electric (TE) and transverse magnetic (TM) polarized light at any angle of incidence [23]. The TPR localization occurs primarily in the non-absorbing part of Tamm structure (within spacer layer) and hence, it is insensitive to dissipative losses in metal layer. The TPR is a suitable alternative to conventional surface plasmons in a wide range of applications such as extraction of single photon emission [9], low threshold lasers [24], sensors [25], and optical switches [26]. In many quantum applications like quantum communication, quantum entanglement, and in solid-state qubits using NV- center, the ZPL emission enhancement with PSB suppression is an essential requirement [12, 13]. The TPR based enhancement of the emission can help in increasing the brightness and emission rate at ZPL relative to phonon mediated broadband emission. The increase in emission brightness and rate is a collective consequence of LDOS enhancement which depends on the emitter's position within cavity and orientation of emitter's dipole moment [27].

Here, we discuss the coupling of single NV- center with an asymmetric Tamm structure to enhance its emission rate and brightness at ZPL wavelength of 640 nm. The proposed easy-to-fabricate asymmetric Tamm structure generates dual TPR at normal-incidence, with first resonance appearing at ZPL of NV- center offering a high $Q$-factor compared to conventional Tamm structure. The enhancement in the decay rate of NV- center and far-field light collection efficiency is studied using finite difference time domain and Green's function method, respectively. We discuss Purcell and light collection efficiency enhancement, from the coupled NV- center, at 640 nm compared to NV- center in a homogeneous dielectric slab.

## II. RESULTS AND DISCUSSIONS

Figure 1(a) represents the schematic of the proposed asymmetric Tamm structure made up of two separate sets of DBRs (DBR$_1$ and DBR$_2$) consisting of different numbers of bi-layers made of titanium dioxide (TiO$_2$) and silicon dioxide (SiO$_2$). The optical thickness of dielectric layers is set according to: $\lambda_o = 2(n_a d_a + n_b d_b)$, where $\lambda_o$ is the central wavelength of DBR's stop-gap, $n_a, n_b$ are the refractive index values of TiO$_2$ and SiO$_2$ layers having thickness $d_a$ and $d_b$, respectively. The $d_a$ value is initially fixed at 60 nm and the calculated value of $d_b$ is ~110 nm for $\lambda_o$ = 640 nm which corresponds to the NV- center ZPL wavelength. The DBR$_1$ is placed above the silver (Ag) layer in such a way that the Ag layer of thickness $d_m$ = 40 nm is sandwiched between DBR$_1$ and DBR$_2$, having different periods. The layers adjacent to Ag layer on either side are taken as two spacer layers with thicknesses $d_{s1}$ and $d_{s2}$. The TPR resonance is very susceptible to $d_a$ or $d_b$ values and for the feasibility of structure, these values are approximated to an integral multiple of 5 nm.

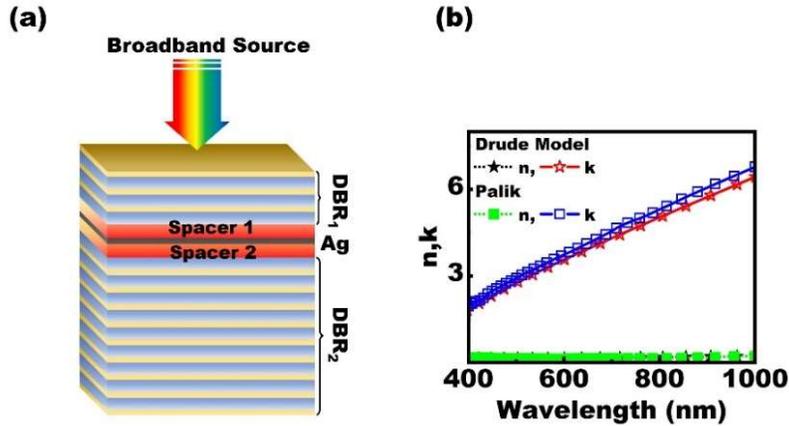

**Figure 1**. (a) Schematic of the asymmetric Tamm structure consists of two DBRs comprised of TiO$_2$ (yellow) and SiO$_2$ (light blue) bilayer and a spacer layer (red); the spacer layers are separated by a thin Ag layer (black) with $d_m$ = 40 nm. (b) The complex refractive index ($\tilde{n} = n + ik$) of Ag layer obtained from Drude model (line with star) and from Palik handbook (line with square).

The condition for the appearance of eigenmodes in conventional Tamm structure is given by $r_m r_{BR} e^{2i\phi} = 1$, with $r_m, r_{BR}$ being the reflection coefficients of metal and DBR and $\phi$ is the phase change across spacer layer thickness. The noble metal Ag follows the behaviour of a free-electron model described by Drude model [28]. The frequency-dependent Ag dielectric function is given by $\varepsilon(\omega) = 1 - \omega_p^2/{\omega^2 + i\Gamma\omega}$, $\Gamma$ is the plasma collision rate related to absorption in Ag layer. In the limit $\omega \ll \omega_p$; for small absorption, the eigen frequency $\omega$ with $n_a > n_b$ can be approximated as [23, 29]:

$$\omega = \frac{\omega_o}{1 + \frac{2\omega_o}{\pi\omega_p}|n_a - n_b|} \tag{1}$$

where $\omega_o$ is the central frequency of the DBR stop gap and $\omega_p$ is bulk plasma frequency of Ag layer. The Drude model offers analytical values of Ag refractive index which are comparable to the index values given in Palik handbook, as shown in Fig. 1(b) [30]. It shows the comparison between analytical values of refractive index ($\tilde{n}$) with real ($n$) and imaginary ($k$) part obtained using Drude model (line with stars) and compared with those values given in Palik handbook (line with square). The dotted and solid line with star, respectively, shows the $n$ and $k$ values of $\tilde{n}$ obtained from the Drude model. The dotted line with squares shows $n$ and $k$ values given by Palik handbook. In the region of our interest (400 nm to 1000 nm), both the Drude model and Palik handbook show nearly same values, and hence index profile given by Palik handbook is used in the following calculations.

### A. The dual Tamm plasmon resonance

The TPR states form at the interface between metal and DBR, while the cavity mode in a conventional planar photonic structure forms deep within the DBR. The strong dip in reflectivity value within the DBR reflectivity spectra signifies the excitation of TPR states and/or photonic cavity states [31]. Hence, the modes can be distinguished based on field intensity localization within the structure. The cavity mode localize in the cavity layer embedded deep inside the DBR, while the TPR mode localize the light at the interface between DBR and metal film [23]. In our calculation, a broadband source emitting light in the wavelength range from 400 to 1000 nm is incident from the top to the DBR$_1$, as shown in Fig. 1(a). The normal-incidence reflectivity spectra is calculated using transfer matrix method (TMM), which involves calculation of incident and reflected field amplitudes at each interface of the structure [32]. The plane wave excitation creates the TPR (black symbols) within the asymmetric Tamm structure, which appears as a dip in the reflectivity spectra of the DBR stop gap (solid red line), as seen in Fig. 2(a). When the Ag layer is absent, an asymmetric cavity mode is manifested at 570 nm (dotted vertical line) for a cavity thickness $d_{s1} + d_{s2} = 95$ nm. When light enters inside the cavity, it gets reflected back and forth between two DBRs. The number of bilayers in DBR$_1$ is less in comparison to the number of layers in DBR$_2$. Thus, 100% light is reflected by the DBR$_2$ whereas the DBR$_1$ reflect minimal light into the cavity layer which result in a shallow reflectivity dip with very low $Q$ value. Here, the confinement is relatively weak, and most of the incident light is reflected toward incident medium. The weak confinement is due to asymmetric nature of photonic cavity with a shallow reflectivity dip at 570 nm.

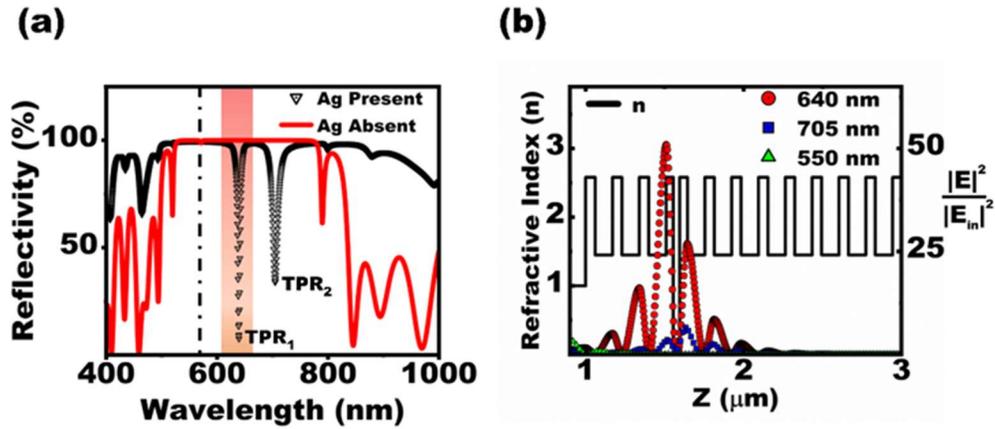

**Figure 2:** (a) Reflectivity spectra of the asymmetric Tamm structure with (symbols) and without (line) Ag layer. (b) The electric field intensity profile within the structure at TPR$_1$ (640 nm) (circle), TPR$_2$ (700 nm) (square), and at off-resonance (550 nm) (triangle) wavelength. The solid line shows the refractive index profile along the growth direction.

The inclusion of Ag layer in-between the two spacer layers ($d_{s1}$ and $d_{s2}$) forms asymmetric Tamm structure. In contrast to the single reflectivity dip in conventional Tamm structure, we observe two spectrally separated reflectivity dips for asymmetric Tamm structure, as shown in Fig. 2(a). The reflectivity dip at lower wavelength is labelled as TPR$_1$ while the higher wavelength reflectivity dip is termed as TPR$_2$. In our calculation, we use Ag layer thickness ($d_m$) of 40 nm, the spacer layer thicknesses of $d_{s1} = 45$ nm and $d_{s2} = 50$ nm to tune TPR$_1$ to 640 nm and TPR$_2$ to 700 nm. It is clear that the coupled TPRs originate when the Ag layer is introduced within the structure, which alters the electric field distribution along growth direction within the structure. Similar hybrid states are also proposed by Kaliteevski *et al*. [33], where TPRs are coupled to the exciton-polaritons modes.

Due to the periodic variation of refractive index in DBR and intrinsic negative dielectric constant of Ag, both DBRs and Ag layer behave as high reflecting mirrors. The peak reflectivity wavelength of DBR is tuned to overlap with the reflectivity spectra of Ag mirror by engineering DBR structural parameters. Accordingly, we optimize structural parameters to obtain field localization at the TPR wavelength in comparison to an off-resonance wavelength, as shown in Fig. 2(b). The refractive index profile (solid line) of asymmetric Tamm structure is presented in Fig. 2(b) for better visualization of the structure. For TPR$_1$ at 640 nm, the field intensity is more confined at the interface between spacer-1 and Ag layer. The corresponding field intensity at 640 nm is estimated to be 55 times (red circles) in comparison to the incident field intensity. At the TPR$_2$ wavelength of 700 nm, the field confinement (blue squares) is achieved between spacer-2 and Ag interface with a low enhancement of 10. The *Q*-factor is higher for TPR$_1$ in comparison to TPR$_2$ and hence better field confinement at 640 nm. Further, the field intensity at an off-resonance wavelength of 550 nm (green triangles) is mostly back-reflected into incident medium and hence, does not show any significant intensity enhancement inside the structure. Hence, we emphasize that only TPR wavelengths mitigate substantial field intensity localization whereas, for off-resonant wavelength, the asymmetric Tamm structure acts as a perfect reflector.

### B. The metal and spacer thickness-dependent Tamm mode

We used optimized values of $d_a$, $d_b$, $d_m$, $d_{s1}$ and $d_{s2}$ to obtain dual TPR with one of the resonances at 640 nm. However, the appropriate choices of these values would enable the achievement of the TPR at any desired wavelength [31]. The changes in the layer thickness values modify the phase relations and thus induce the shift in TPR. As we have seen earlier when $d_m$ value is set to zero with $d_{s1}$ = 45 nm and $d_{s2}$ = 50 nm, the proposed structure incorporates an asymmetric photonic cavity with a cavity thickness of 95 nm, as shown in Fig. 2(a) (solid red line). The emergence of two resonances (TPR[1] and TPR[2]) depends on $d_m$ value and with an increase in $d_m$ value, the resonances start appearing from lower and higher spectral regions, as shown in Fig. 3(a). It is observed that increase in $d_m$ value redshifts the TPR[1] while the TPR[2] is blue-shifted within DBR stop gap. Thus, both TPR[1] and TPR[2] evolve and move towards each other as $d_m$ value is increased.

Moreover, it is seen that both resonances are halted near a critical $d_m$ value of 65 nm, which is due to the phase shift becoming almost constant at higher $d_m$ values [34]. [34]. Hence, TPR is insensitive to a further increase in $d_m$ values [35]. For $d_m$ > 65 nm, the TPR[2] starts disappearing as light is unable to propagate to other side (DBR[2]) of asymmetric Tamm structure due to significant light absorption induced by Ag layer. This makes asymmetric Tamm structure behave like a conventional Tamm structure with a single Tamm mode at 640 nm.

Figure 3(b) shows the contourmap obtained using the calculated reflectivity spectra at various $d_{s1}$ values while $d_{s2}$ and $d_m$ are fixed at 50 nm and 40 nm, respectively. When $d_{s1}$ = 0 nm, the adjacent layer on top of Ag layer becomes quarter-wave thick with an index $n_b$ associated with DBR[1]. Therefore, it induces a phase change of $\pi$ into the incoming light and hence DBR[1] does not contribute anymore to the formation of dual TPR modes and the structure becomes a conventional Tamm structure. Thus, we naively expect a single TPR resonance originated due to Ag layer and DBR[2] in the reflectivity spectra as seen at 686 nm ($d_{s1}$ = 0 nm) in Fig. 3(b). When the value of $d_{s1}$ increases, a new resonant mode emerges from the short-wavelength side with a long-wavelength shift. With an increase in the $d_{s1}$ value, an overall phase is acquired that shifts the mode to long-wavelength side. Interestingly, both modes exhibit an avoided crossing for $d_{s1}$ = 52 nm, as expected from the anti-crossing nature of the mode resonances. When the $d_{s1}$ value is inadequate, the coupling between TPR$_1$ and TPR$_2$ is weak. In this weak coupling regime, the TPR resonances in reflectivity spectra appear near the resonant mode exhibited by the conventional Tamm structure with two different values of $d_{s1}$ and $d_{s2}$. The higher value of $d_{s1}$ enhances the coupling between TPR$_1$ and TPR$_2$ which forms a hybrid state characteristically distinct from single TPR.

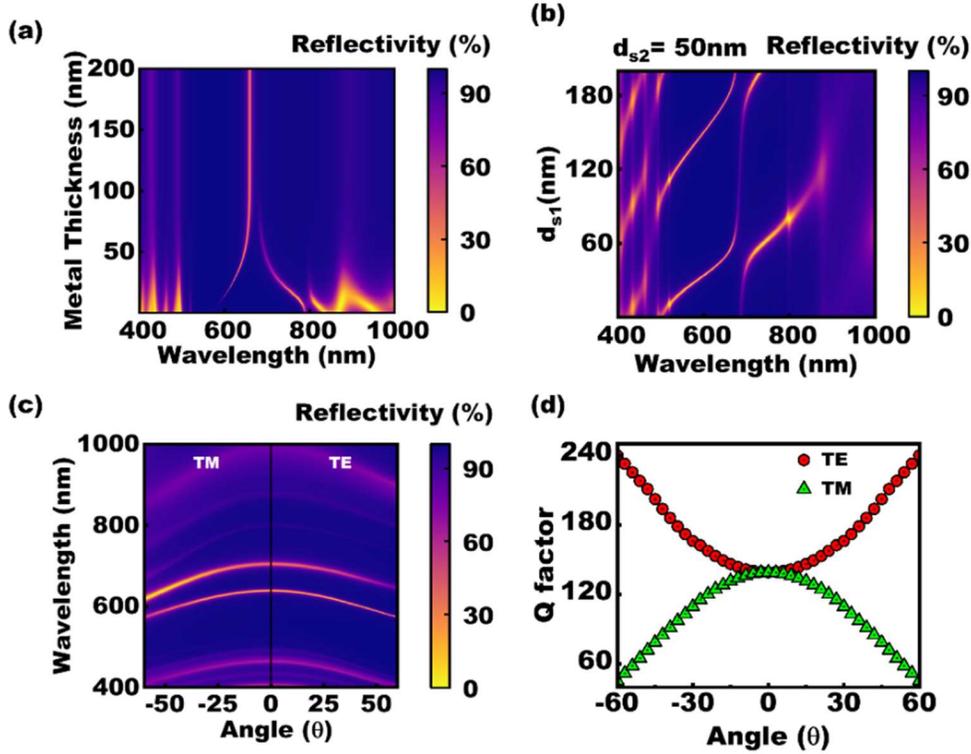

**Figure 3:** (a) Variation of TPR wavelength as a function of $d_m$ values $d_{s1} = 45$ nm and $d_{s2} = 50$ nm. (b) Reflectivity value for various $d_{s1}$ values with $d_{s2} = 50$ nm. At 680 nm, the TPR$_1$ and TPR$_2$ shows an avoided crossing of modes. (c) The angle-dependent variation of TPR$_1$ and TPR$_2$ for TE and TM polarization. (d) The variation of $Q$-factor at TPR$_1$ as a function of angle of incidence for TE and TM polarization.

As we have seen that the TPR can be excited using both TE and TM polarization as both polarization modes exist within light cone. Figure 3(c) shows the angular dispersion of TPR mode in the asymmetric Tamm structure. The negative angles correspond to TM polarization while the positive angles correspond to TE polarization. We have used structural parameters such as $d_{s1} = 45$ nm, $d_{s2} = 50$ nm and $d_m = 40$ nm to achieve TPR$_1$ at 640 nm for normal-incidence. The dual modes show angular dependence similar to conventional Tamm mode. The spectral width of TPR$_1$ is relatively low compared to TPR$_2$ and hence higher $Q$-factor for TPR$_1$. The $Q$-factor is calculated as $\lambda/\Delta\lambda$, where $\Delta\lambda$ is the full width at half maximum (FWHM) of the reflectivity dip at TPR wavelength $\lambda$. The $\Delta\lambda$ at normal-incidence for TPR$_1$ is 4.6 nm which is quite less in comparison to $\Delta\lambda = 13.8$ nm of conventional Tamm structure at the same TPR wavelength. Figure 3(d) shows the variation of $Q$-factor with angle of incidence for the asymmetric Tamm structure. The $Q$-factor is estimated to be 140 at normal incidence for TPR$_1$, and it increases (decreases) for TE (TM) polarization as the angle of incidence increases seen in Fig. 3(d). The $Q$-factor for TPR mode in a conventional Tamm structure is ~45 at a TPR wavelength of 640 nm. Thus the proposed asymmetric Tamm structure provides high $Q$ modes that enable significant modulation of single quantum emitter emission properties.

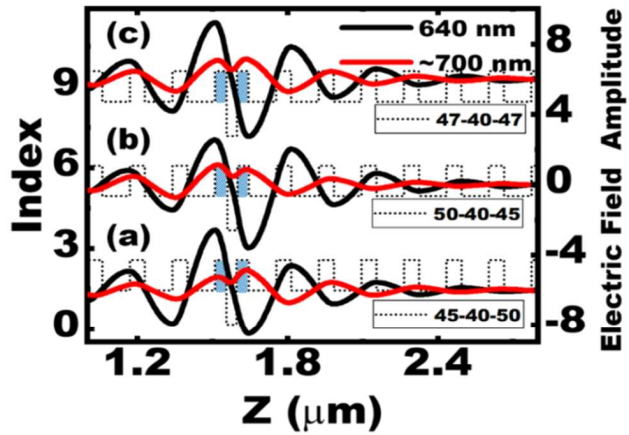

| Figure | $d_{s1}$ (nm) | $d_{s2}$ (nm) | Index offset; Multiplier | Field offset; Multiplier |
|---|---|---|---|---|
| (a) | 45 | 50 | 0; 1 | -6; 0.5 |
| (b) | 50 | 45 | 1; 3.5 | 0; 0.5 |
| (c) | 47 | 47 | 1; 7 | 6; 0.5 |

**Figure 4:** The electric field amplitude within the structure for two different TPR modes at 640 nm (black line) and 700nm (red line). The field is incident from left to right. The structural parameters along with offset and multipliers are shown in the table. In all cases $d_m = 40$ nm and refractive index profile is shown by dotted curves. The blue patterned area highlights the spacer layer.

As we have seen earlier, the asymmetric Tamm structure supports resonances with field intensity localization at the upper or the lower spacer layer-Ag interface. Depending upon specific structural parameters, we can tune the spatial confinement of field intensity and there would be three different cases of field confinement: (a) strong field confinement at spacer-1-Ag interface, (b) strong field confinement at spacer-2-Ag interface, and (c) an equalized field confinement at interface between both spacer layers and Ag layer. The light is incident at normal-incidence on the structure along z-direction and the corresponding index profile (dotted line) is shown in Fig. 4. The electric field profiles at TPR$_1$(640 nm) and TPR$_2$ (700 nm) show different propagation characteristics within structure. The TPR$_1$ changes sign within Ag layer while TPR$_2$ do not change the sign but attains a minimum value. The field confinement associated with TPR$_1$ resonance is not evanescent even in the presence of the Ag layer. This indicates that there would be a reduction in absorption at 640 nm, which results in a high Q-factor for TPR$_1$. The NV-center coupled with such a resonant mode, helps in radiating higher fields out of the structure as the absorption gets mitigated.

Figure 4(a) shows the field intensity variation through the asymmetric Tamm structure with $d_m = 40$ nm, $d_{s1} = 45$ nm, and $d_{s2} = 50$ nm. It is seen that the field confinement is more prominent at the interface between spacer-1 and Ag at TPR$_1$ of 640 nm (black solid line). However, the field confinement is less prominent and appears at the interface between spacer-2 and Ag for TPR$_2$ (red solid line). Figure 5(b) depicts the field intensity variation for $d_m = 40$ nm, $d_{s1} = 50$ nm, and $d_{s2} = 45$ nm, with prominent field confinement achieved at the interface between spacer-2 and Ag for TPR$_1$. However, minimal confinement occurs at the interface between spacer-1 and Ag for TPR$_2$. In the case of equal spacer thickness ($d_{s1} = d_{s2} = 47$ nm), nearly the same field confinement is obtained at both interfaces for TPR$_1$. However, the field confinement is less prominent, with almost equal strength for TPR$_2$. Thus varying the thickness of $d_{s1}$ and $d_{s2}$, we can tune the field confinement at the TPR, which appears at one of the interfaces or at both interfaces simultaneously, as shown in Fig. 4. Moreover, the fields corresponding to TPR$_1$ and TPR$_2$ are in the phase before they reach Ag layer. After crossing the Ag layer, the fields become out of phase in all the above three cases, as shown in Fig. 4. This eventually indicates that both resonances could show an avoided-crossing characteristic at a particular wavelength. The out-of-phase wavelength obtained from the field analysis appears at 680 nm, which matches with avoided crossing wavelength in the TPR modes, shown in Fig. 3(b). At 680 nm, the field amplitudes of both resonances cancel each other and hence cannot propagate inside the structure that results in higher reflectivity, as seen in Fig. 3(b). The TPR$_1$ wavelength is tuned in such a way that the asymmetric Tamm structure can be used to enhance the NV- center decay rate at ZPL wavelength. In addition, TPR$_1$ exhibit a higher $Q$-factor and hence higher light collection efficiency (LCE) can be achieved at 640 nm as discussed in the following.

### C. Spontaneous decay rate enhancement

According to Fermi's Golden Rule, the spontaneous emission decay rate ($\gamma$) of an excited quantum emitter depends on the available density of optical states (DOS) at the emission wavelength, which states that [6]:

$$\gamma = \frac{2\pi}{\hbar^2}\Sigma_f |< f|\widehat{H}_I|i >|^2 \delta(\omega_f - \omega_i) \qquad (2)$$

Here $\widehat{H}_I = -\hat{\mu}.\widehat{E}$ is the interaction Hamiltonian in the dipole approximation in terms of dipole moment operator ($\hat{\mu}$) and electric field operator ($\widehat{E}$). The quantum emitter is approximated as an oscillating electric dipole which acts as a source of electromagnetic radiation. The decay rate of a dipolar emitter depends on its orientation and the local environment surrounding it through DOS [4, 6]. As the emission from dipolar emitters is sensitive to orientation and local environment, we find the partial local density of optical states (PLDOS) is a more relevant quantity to calculate the orientation-dependent decay rate ($\gamma_\mu$). A dipole aligned along unit vector $\hat{n}_\mu$ will interact strongly with those modes for which the electric field is polarized in the same direction. The interaction is expected to be weak for perpendicularly polarized modes. The $\gamma_p$ for a dipole in free space is written in terms of PLDOS as [6]:

$$\gamma_p = \frac{2\omega_o}{3\hbar\varepsilon_o}|\vec{\mu}|^2 3\sum_k[\hat{n}_\mu\cdot(\vec{u}_k\vec{u}_k^*)\cdot\hat{n}_\mu]\delta(\omega_k-\omega_o) \qquad (3)$$

$$\gamma_p = \frac{2\omega_o}{3\hbar\varepsilon_o}|\vec{\mu}|^2 \rho_p(\vec{r}_o,\omega_o) \qquad (4)$$

where $\rho_p(\vec{r}_o,\omega_o)$ is PLDOS which is sensitive to dipole orientation. The $\vec{r}_o$ corresponds to the origin of dipole which emits at a frequency $\omega_o$. The $\hat{n}_\mu$ is unit vector in the direction of dipole moment $\vec{\mu}$, $\vec{u}_k$ are the energy of normal modes, the sum over $\vec{k}$ refers to summation over all modes, and $\omega_k$ denotes the mode frequency $\vec{k}$. The $\delta(\omega_k-\omega_o)$ accounts for the frequencies of available optical states and with significant dissipation, the delta function loses its meaning in eq. (3). The mode amplitudes decrease over time due to dissipation and the eigen frequencies for such system must be complex with null real eigen frequencies. It is more convenient to represent PLDOS in Green's function formalism, which is a powerful tool to calculate PLDOS and hence spontaneous emission decay rate [6]. The electromagnetic Green's function is a dyadic quantity that estimates the radiated field due to a point source located at $\vec{r}_o$. To consider all components of the source, we need the Green's function as a tensor quantity denoted as $\overleftrightarrow{G}$. It is a compact notation for the three components of Green's function corresponding to three distinct dipole orientations along *x-y-z*-axis. The $\rho_p$ is related to $\overleftrightarrow{G}$ by the relation [6]:

$$\rho_p(\vec{r}_o,\omega_o) = \frac{6\omega_o}{\pi c^2}[\hat{n}_\mu\cdot\text{Im}\{\overleftrightarrow{G}(\vec{r}_o,\vec{r}_o;\omega_o)\}\cdot\hat{n}_\mu] \qquad (5)$$

Here, the $\rho_p$ represents PLDOS when the dipole is placed in air/vacuum. For other dielectric mediums, the eq. (5) needs to be modified to take into account the refractive index contribution. Moreover, we also need to calculate the LDOS for the emitter, which emits in all spatial directions. The LDOS is the average of PLDOS over orthogonal dipole orientations along *x-y-z*-axis. The spontaneous decay rate $\gamma_l$ is related to LDOS in terms of $\overleftrightarrow{G}$ as:

$$\gamma_l = \frac{2\omega_o}{3\hbar\varepsilon_o}|\vec{\mu}|^2 \rho_l(\vec{r}_o,\omega_o) \qquad (6)$$

$$\gamma_l = \frac{2\omega_o}{3\hbar\varepsilon_o}|\vec{\mu}|^2 \frac{2\omega_o}{\pi c^2} \operatorname{Im}\{\operatorname{Tr}[\vec{G}(\vec{r}_o,\vec{r}_o;\omega_o)]\} \tag{7}$$

where $\rho_l(\vec{r}_o,\omega_o) = \frac{2\omega_o}{\pi c^2} \operatorname{Im}\{\operatorname{Tr}[\vec{G}(\vec{r}_o,\vec{r}_o;\omega_o)]\}$ is the LDOS and $\operatorname{Tr}[\vec{G}(\vec{r}_o,\vec{r}_o;\omega_o)]$ denotes the trace of dyadic 3x3 tensor matrix $\vec{G}$. The value of LDOS determines how quickly an excited atom decays to its ground state, producing a photon in the process. Thus we need the information of $\vec{G}$ within a system in which the emitter is emitting the radiation.

The LDOS calculations are performed using a three-dimensional (3D) finite-difference time-domain method using Ansys Lumerical. A single NV- center emitting at ZPL wavelength of 640 nm is positioned within spacer-1 layer, wherein maximum electric field intensity is obtained for TPR$_1$, as seen in Fig. 2(b). The simulation is performed within a 2.25 $\mu m^2$ region in *x-y* plane with a mesh size of 4 nm using perfectly matching boundary conditions in *x-y-z* directions. The resultant radiated power is detected by a 3D transmission box that calculates net outward power flow from dipole within simulation region. The resulting electric field at dipole location within spacer-1 computes Green's dyadic $\vec{G}$:

$$\vec{E}(\vec{r}) = \frac{\omega^2 \mu_o \mu_r}{n^2} \vec{G}(\vec{r},\vec{r}_o) \cdot \vec{\mu} \tag{8}$$

where the electric field $\vec{E}$ is calculated at the position $\vec{r}$, $n = \sqrt{\mu_r \varepsilon_r}$ is the refractive index of material with relative permeability and permittivity as $\mu_r$ and $\varepsilon_r$, respectively. The dipole is placed within spacer-1 layer which has an index of $n_a$.

### D. Purcell enhancement and light collection efficiency

The interaction of single NV- center with an enhanced localized field at TPR$_1$ is analyzed using the Purcell factor calculations. When local environment is not homogeneous, eq. (3) suggest that $\gamma$ is discrete for parallel and perpendicular orientations of emitter. The Purcell enhancement is defined as: $\frac{\gamma}{\gamma_o}$, where $\gamma_o$ and $\gamma$ are the spontaneous decay rate when NV- center is embedded in a dielectric slab and when NV- center is in asymmetric Tamm structure, respectively. The Purcell enhancement is averaged for emitters oriented in parallel and perpendicular directions with respect to the plane of the structure (*x-y* plane) to account for all distinct emitter orientations. The emission properties of the NV- center are studied using radiating point dipole source. The dipole is positioned in spacer-1 at a distance of 30 nm away from Ag layer, as shown in Fig. 5(a). If the separation distance < 30 nm, there is a sharp rise in DOS due to surface plasmon modes excited at Ag-TiO$_2$ interface. However, these states include both radiative and non-radiative modes induced by the environment. At a smaller distance, the non-radiative decay channels dominate which results in a significant emission quenching. Thus positioning the emitter at a distance of 30 nm from Ag layer in spacer-1 layer ensures that dipole emission quenching due to excited plasmon mode is avoided. Fig. 5(a) shows the electric field intensity of the emitter positioned in the dielectric spacer-1 layer and emitting at 640 nm. The dashed lines in Fig. 5(a) separate the SiO$_2$-TiO$_2$ layers, and grey area at $Z = 0$ nm encloses the Ag layer. We observe localized field intensity at the interface between spacer-1 and Ag layers due to field confinement induced by TPR$_1$. Moreover, the field decays off along z-axis with high intensity in the SiO$_2$ and low intensity in TiO$_2$ layers. The far-field emission pattern at the resonance wavelength of 640 nm and at different off-resonance wavelengths is shown in Fig. 5(b). The on-resonance emission pattern shows four times emission intensity at the far field compared to off-resonance wavelength, clearly indicates the off-resonant emission intensity suppression.

The asymmetric Tamm structure shows dual TPR in the reflectivity spectra with distinct field localization in the spacer layer, as seen in Fig. 2(a). The resonance wavelength has a strong dependence on the $d_m$ value, as seen in Fig. 3(a) and thus, we should expect Purcell enhancement also depends on $d_m$ value at different TPR wavelengths. Figure 5(c) shows the Purcell enhancement factors, estimated for different $d_m$ values. For $d_m \geq 40$ nm, we observe Purcell enhancement for both modes, with TPR$_1$ mode shows a higher Purcell enhancement factor of 4 in contrast to TPR$_2$ mode, as shown in Fig. 5(c). Moreover, it is evident that Purcell enhancement factor is comparatively low for smaller $d_m$ values at the TPR$_1$. This is due to the fact that thin Ag film, below a certain $d_m$ value, is unable to sustain TPR mode and thus results in weak emitter-TPR coupling. The wavelength-dependent Purcell enhancement averaged over all dipole orientations (parallel and perpendicular) shows a strong dependence on the $d_m$ value. With an increase in its value, the Purcell factor increases and saturates to 4 for $d_m \geq 40$ nm. It should be noted that the Purcell enhancement peak shows a redshift with an increase in $d_m$ value, which follows from Fig. 3(a). The presence of a thick Ag layer (high $d_m$ value) increases absorption, thus inhibiting an effective field that can escape the structure. However, a minimum $d_m$ value is necessary to sustain the TPR mode. Thus, it is a trade-off between the required Purcell enhancement and the amount of light escaping from the structure through the judicious choice of $d_m$ value. The optimum $d_m$ value is 40 nm, corresponding to high Purcell factor at TPR$_1$ which helps to acquire fast emitted photons from the asymmetric Tamm structure.

The calculated wavelength-dependent Purcell enhancement shows a lower value for shorter wavelengths and increases to its maximum value at 640 nm for $d_m = 40$ nm and have a sudden drop after 640 nm. This enhancement is a result of excitations of both surface plasmon polaritons and/or TPR within the structure. The emitted light from single NV- the center is coupled to the evanescent field and surface plasmon polaritons are excited accordingly. However, the TPR excitations are primarily dominated at higher wavelength regions. It is well known that the evanescent field penetration is proportional to the wavelength [6]. Thus, in the shorter wavelength regime (< 450 nm), the evanescent field penetration is minimal for the dipole placed at a distance of 30 nm above the Ag layer. This results in reduced near-field coupling and thus an insignificant Purcell enhancement factor. However, for the spectral region between 450 to 500 nm, the increase in the Purcell enhancement factor is mainly due to emitter coupling with surface plasmon polaritons. For the wavelength above 570 nm, the Purcell enhancement is primarily due to emitter coupling with TPRs, which can be understood using the parabolic dispersion of TPR [23].

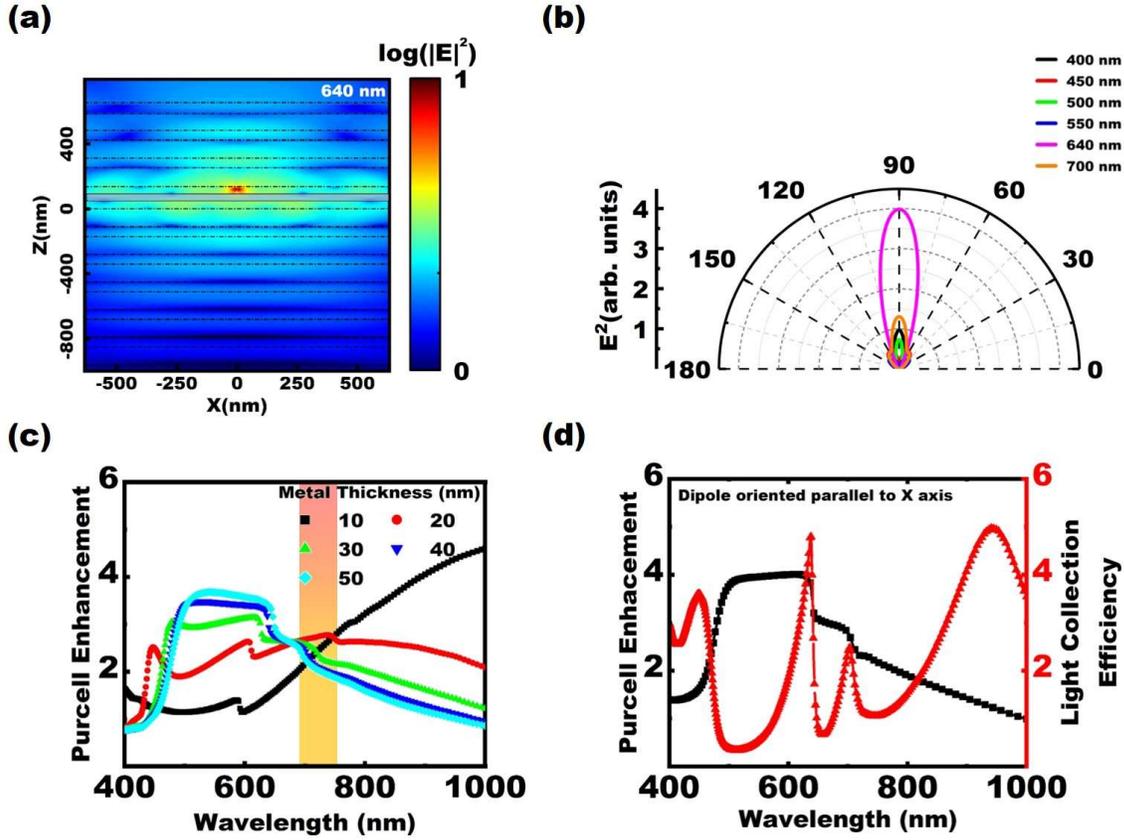

**Figure 5:** (a) The electric field intensity variation associated with an emitted field for the emitter placed in the spacer layer. The high magnitude region (red region above Ag layer) is the location where dipole is placed. (b) Far-field patterns of light emission intensity at various wavelengths (c) The wavelength-dependent Purcell enhancement for NV- center located in the spacer-1 layer. The shaded region corresponds to TPR$_1$ with maximum Purcell enhancement of 4. (d) Purcell enhancement and light collection efficiency at various wavelengths for the dipole oriented parallel to structure with $d_m = 40$ nm.

The $\gamma$ can also be given in terms of group velocity dispersion ($\frac{d\omega_k}{dk}$) such that $\gamma \propto \frac{d(k^2)}{d\omega_k}$ and hence the Purcell factor [36]. In the spectral range between 570 nm to 640 nm, $\frac{d\omega}{dk}$ decreases gradually that result in an increase in Purcell enhancement (dipole oriented parallel to the structure). At 640 nm, the $\frac{d\omega}{dk}$ reaches its minimum value and accordingly, the maximum Purcell enhancement of 4 is obtained. There is a steep slope above 640 nm, which suggest a decrease in coupling strength above 640 nm, which follows from angle-dependent reflectivity contour map shown in Fig. 3(c). This emphasizes that the high Purcell enhancement at TPR wavelength is due to emitter coupling with TPR and above TPR$_1$, the Purcell enhancement factor decreases rapidly. We have taken the average Purcell enhancement that includes both kinds of dipole orientations. However, the dipole would interact strongly with those modes whose electric field is polarized in the same direction as the dipole orientation. Thus the main contribution to Purcell enhancement shown in Fig. 5(c) arises from the parallel-oriented dipole. In addition to the Purcell enhancement, LCE is an important parameter to be considered for quantum emitter trapped in photonic cavities [37]. The LCE quantifies the amount of light collected at the far-field using a suitable detector from single emitter coupled to a photonic structure. Even if the emission rate or Purcell enhancement is high, the LCE is also required to be high to achieve maximum collection of fast photons. Figure 5(d) shows Purcell enhancement (black circle) and LCE (red triangle) for parallel-oriented dipole as a function of wavelength. The emitted light is collected in the far-field within a cone of half-angle $60^0$ above the structure to estimate the LCE. This angular range corresponds to a microscopic objective of numerical aperture 0.85, typically used in experiments involving single quantum emitters. The LCE is estimated as LCE = $\frac{FFC_{60}}{P_e}$, where $FFC_{60}$ is the far field enhancement of emission intensity collected within $60^0$ cone above the metal layer. It is obtained as the ratio between emission intensity collected within a light cone of $60^0$ from asymmetric Tamm structure to the emission from a dielectric slab (TiO$_2$). The P$_e$ is the Purcell enhancement for the dipole aligned parallel to the structure. The results clearly show a large LCE for the emitted light corresponding to TPR$_1$. It manifests that the TPR play a significant role in extracting the emitted photons in comparison to light collection using surface plasmon polaritons. The NV- center coupled to an asymmetric Tamm structure provides maximum Purcell enhancement along with higher LCE at 640 nm. The structure shows the maximum LCE factor of 5 and 2.5 at 640 nm and 700 nm, respectively. The light collection is improved 5 times in comparison to the emitter emitting in a dielectric slab at 640 nm. It is found that after 640 nm, there is a sharp decrease in LCE compared to the Purcell enhancement factor. The drop in Purcell enhancement is due to the absence of TPR above 640 nm (Fig. 3(b)), and remaining enhancement is induced by surface plasmon polaritons. Moreover, for the LCE, only TPRs excitations provide the maximum extraction of emitted photons, and the absence of TPRs causes a sudden decrease in LCE. Further, the $Q$-factor of $TPR_1$ is high, which also helps in cutting emissions above 640 nm. Thus, for NV- center, the PSB emission near ZPL can be efficiently suppressed, providing an enhanced ZPL emission at 640 nm. The coupling of NV- center with TPR in the asymmetric Tamm structure ensures an enhanced spontaneous emission decay rate at 640 nm. Furthermore, the high $Q$-factor and low dissipative losses give high collection efficiency of the emitted photons. The resonances can be controlled by varying the thickness of the metal layer and the adjacent dielectric (spacer) layer. Thus, the structure can be tuned to enhance the emission from other quantum emitters like Si Vacancy center, hBN integrated with asymmetric Tamm structures.

## III. CONCLUSIONS

The asymmetric Tamm structure offers dual resonances with one of them having high quality factor which can be used to tune the emission rate of a single NV- center. The structure offers high-quality resonance with $\Delta\lambda = 4.6$ nm in comparison to conventional Tamm structure ($\Delta\lambda = 13.8$ nm). The coupling of NV- center, emitting at a ZPL of 640 nm, shows the enhancement in the decay rate up to a factor of 4 compared to NV- center emitting in a dielectric slab. This arises due to strong light confinement at metal-dielectric interfaces induced by the Tamm plasmon resonance mode. We have obtained four times enhanced Purcell factor due to the modification in LDOS induced by Tamm plasmon resonance. The proposed asymmetric Tamm structure helps in getting a five times more light collection at 640 nm in comparison to light collection from NV- center embedded in a dielectric slab. The structure can be fabricated over a large area which is possible using the current sample fabrication tools. The proposed asymmetric Tamm structure offers a paradigm shift in achieving bright single photon emission with enhanced rate with better collection efficiency.

Acknowledgements

The authors would like to acknowledge the financial support from IIT Ropar, DST-RFBR [INT/RUS/RFBR/P-318]; DST-ICPS [DST/ICPS/QuST/Theme-2/2019/General], DST-SERB [SB/SJF/2020-21/05]. RVN acknowledges the Swarnajayanti Fellowship (DST/SJF/PSA-01/2019-20).